\documentclass[twocolumn]{aastex62}
\usepackage{amsmath}
\usepackage{graphicx}
\usepackage{color}

\usepackage{amsmath}
\usepackage{graphicx}

\newcommand{\beq}{\begin{equation}}
\newcommand{\eeq}{\end{equation}}

\begin{document}

\title{Detecting Black Hole Occultations by Stars with Space Interferometric Telescopes}


\author{Pierre Christian}
\affiliation{Astronomy Department, University of Arizona, 933 North Cherry Avenue, Tucson, AZ 85721, USA}
\author{Abraham Loeb} 
\affiliation{Harvard Smithsonian Center for Astrophysics, 60 Garden Street, MS-10, Cambridge, MA 02138, USA}

\begin{abstract}
We show that the occultation of Sagittarius A* by stars can be detected with space-based or space-ground very-long-baseline-interferometers (SVLBIs), with an expected event rate that is high due to relativistic precession. We compute the tell-tale signal of an occultation event, and describe methods to flag non-occultation events that can masquerade as the signal. \\
\\
\end{abstract}

\section{Introduction}

A novel avenue for studying astrophysical black holes on event horizon scales is through high resolution electromagnetic observations \citep{1973blho.conf..215B,1979A&A....75..228L,2000ApJ...528L..13F,2004ApJ...611..996T,2010ApJ...718..446J}. The feasibility of this technique using ground based very-long-baseline-interferometers (VLBIs) has been demonstrated by the Event Horizon Telescope (EHT) \citep{PaperI,PaperII,PaperIII,PaperIV,PaperV,PaperVI}. The first EHT observations of the supermassive black hole at the center of the M87 galaxy (henceforth M87) were conducted at $1.3$mm, with the longest baseline providing a resolution of $\sim25\mu$as \citep{PaperII}. While this is sufficient to resolve the two largest black holes in the sky, M87 and Sagittarius A* (Sgr A*), at scales comparable to their event horizons, the first EHT observations cannot observe phenomena that generate small scale spatial variations on much smaller scales. 

Prior to the $1.3$mm observations, the intrinsic size of the Sgr A* radio source was measured to be $\sim 2$ astronomical units (AU) at $7$mm \citep{2004Sci...304..704B} and $\sim 1$ AU at $3$ mm \citep{2005Natur.438...62S}. While these sizes are dominated by scattering, the wavelength dependence of the intrinsic size implies that measurements at shorter wavelengths have the potential to resolve the source at scales that are comparable to the physical size of the central object. Indeed, a measurement at $1.3$ mm showed structure at scales comparable to the event horizon of a $\sim 4\times 10^6 M_\odot$ black hole \citep{2008Natur.455...78D}. This measurement was followed by VLBI detections of time variability \citep{2011ApJ...727L..36F}, resolved magnetic field structure \citep{2015Sci...350.1242J}, and asymmetry \citep{2016ApJ...820...90F} at $1.3$mm. Another observation at the same wavelength using a larger array with longer baselines also confirmed the detection of a highly compact source \citep{2018ApJ...859...60L}.

There are two main avenues for improving the resolution of a VLBI observatory: by utilizing higher observing frequencies or through the addition of stations that provide longer baselines. On the second front, it is possible to add observing stations in space to the current EHT array, thus upgrading it by the addition of baselines that can in principle be much longer than the Earth's diameter. Another possibility is to perform observations with only stations in space, by combining multiple orbiting satellites. Utilizing their much longer baselines, SVLBIs can operate at longer wavelengths while still improving their resolutions compared to Earth-based observatories. Besides their significantly higher resolution, space-based VLBIs (SVLBIs) have the advantage of not being affected by atmospheric effects. 

The first SVLBI observations were conducted by combining elements of the Tracking and Data Relay Satellite System (TDRSS) with ground-based observatories on Earth \citep{spacevlbi1,spacevlbi2a,spacevlbi2b,spacevlbi3}. This inaugural SVLBI array included a $4.9$m radio antenna on a geostationary orbit as its space-based antenna at $2.3$ and $15$ GHz. The first dedicated SVLBI station was the Highly Advanced Laboratory for Communications and Astronomy (HALCA) satellite as part of the VLBI Space Observatory Programme (VSOP) \citep{spacevlbi4,spacevlbi5}. The HALCA satellite carried an $8$m antenna in an elliptical orbit with an apogee at $28,000$km from the Earth's center and further demonstrated the practical possibility of ground-space SVLBIs. The most recent SVLBI array to be deployed is the RadioAstron project\footnote{http://www.asc.rssi.ru/radioastron/}, which included a $10$m observatory onboard the satellite Spektr-R \citep{spacevlbi6}. The RadioAstron project possessed a maximum baseline of $\sim3.3 \times 10^5$km, and a resolution that is in principle higher than that of the first EHT observations. However, RadioAstron utilizes an observing frequency that is too low to pierce through the plasma surrounding M87 or Sgr A*.

Theoretical studies utilizing synthetic observations of general relativistic simulations show that few satellites on medium Earth orbits can provide resolutions enough to probe small scale spatial structure of the emission structure around Sgr A* as long as the obits of the satellites can be reconstructed accurately \citep{Roelofs}. Further, emission from the vicinity of a black hole include a thin 'photon ring' component consisting of subrings indexed by the number of times that the photons underwent around the black hole \citep{2019PhRvD.100b4018G}. While numerical calculations show that this component produces a strong interferometric signature, the current EHT does not possess the necessary resolution to resolve it, and detection of the individual subrings requires the resolution provided by SVLBIs \citep{Johnsoneaaz1310}.

In this paper, we provide another motivation for an SVLBI observatory. Occultation events, where a dim object covers a portion of the supermassive black hole emission will produce a small scale variation on the black hole emission profile that could be observed by SVLBIs. In Section \ref{sec:possible_oc}, we discuss the population and sizes of stars close to Sgr A* and their role as possible occulters. In Section \ref{sec:precess}, we show that general relativistic precession can greatly increase the transit probabilities of objects in orbit around Sgr A*. In Section \ref{sec:model} we describe our model for the signal of a stellar occultation event. In Section \ref{sec:signal}, we discuss how such a signal is seen by an SVLBI and provide methods to reject false positives using only SVLBI observables. Finally, in Section \ref{sec:conclusion}, we provide our concluding remarks.

\section{Analysis of possible occulters}  \label{sec:possible_oc}

\begin{figure}
\centering
\includegraphics[width=3.3in]{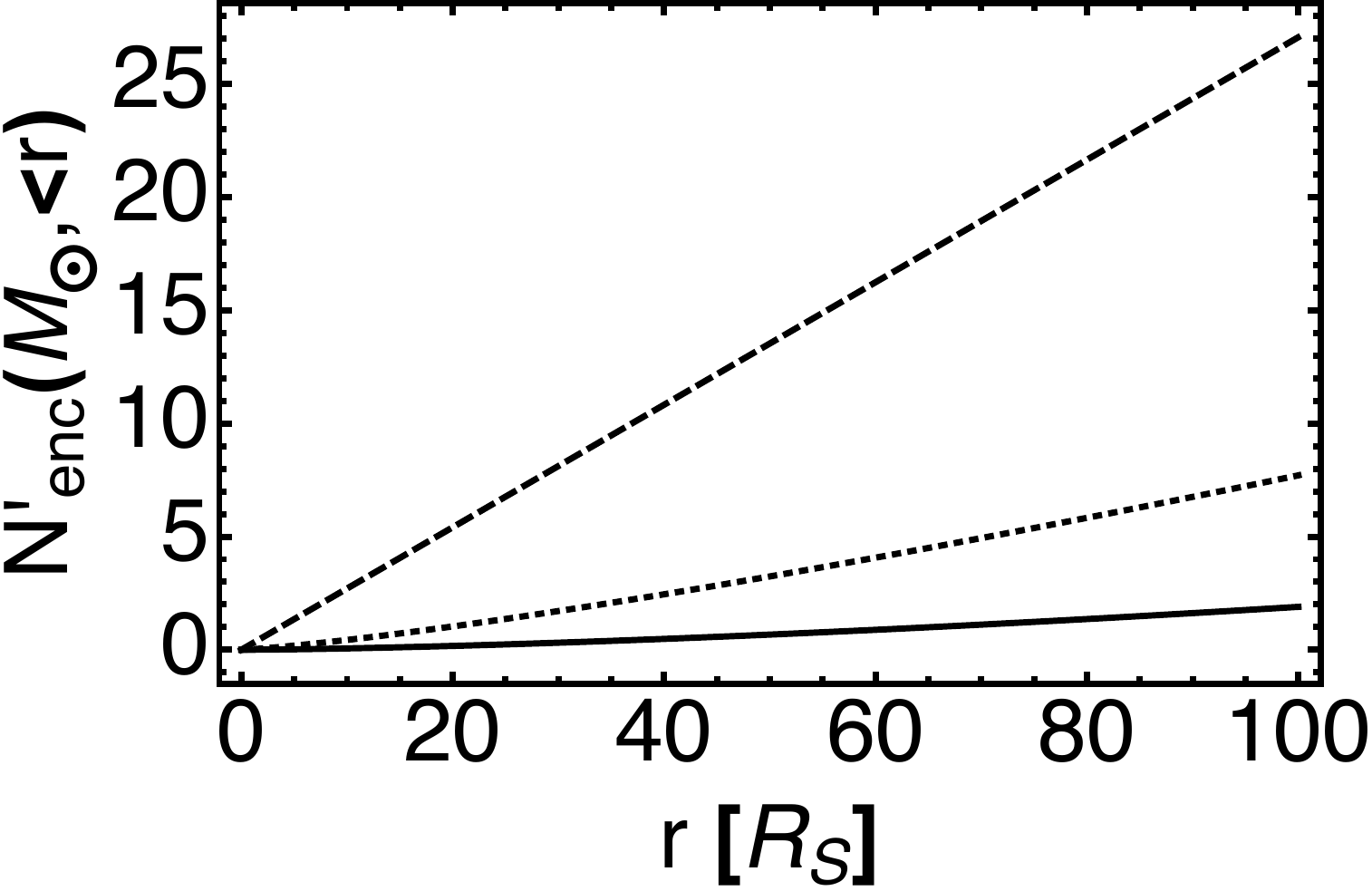} 
\caption{Enclosed number of stars within a logarithmic mass bin centered at $1 M_\odot$ as a function of distance from Sgr A* with $\gamma$ values $1.5$ (solid), $1.75$ (dotted), and $2$ (dashed).}
\label{fig:Nenc}
\end{figure}

\subsection{Population of stellar occulters}

The most probable occulters are stars orbiting close to Sgr A*. In particular, the existence of the S stars, both detected \citep{2002Natur.419..694S,2003ApJ...586L.127G,2005ApJ...620..744G,2005ApJ...628..246E,2009ApJ...692.1075G,2017ApJ...837...30G,2017ApJ...847..120H}, as well as expected but hitherto undetected \citep{2018MNRAS.476.3600W}, suggests the possible existence of a cluster of stars orbiting very close to Sgr A* \citep{Genzel,Sstar1}. While the immense tidal force of Sgr A* prohibits standard star formation mechanisms from forming these stars in situ \citep{1993ApJ...408..496M}, this star cluster can be the formed  through processes such as dynamical interactions of two stellar disks \citep{2008ApJ...683L.151L} or multiple 3-body exchanges between Sgr A* and an in-falling binary \citep{2003ApJ...592..935G}, perhaps through the aid of a massive perturber \citep{2007ApJ...656..709P}. In this section, we will argue that the existence of this star cluster implies an appreciable number of occulters near the tidal disruption radius at $\sim 10 R_S$, where $R_S \approx 10^{12}$cm is the Schwarzschild radius of Sgr A*. 

If we assume the star cluster to be described by the Kroupa mass function \citep{Kroupa}, 
\beq 
\mathrm{N}' \equiv \frac{d \mathrm{N}}{d \log m} \propto m^{-1.3} \; ,
\eeq
where $m$ is the stellar mass and $\mathrm{N} dM$ is the number of stars in the mass range $m$ to $m+dm$, then we can relate the number of stars enclosed within radius $r$ from Sgr A* of an arbitrary mass $M$ to the number of stars with mass $10 M_\odot$ per logarithmic mass bin,
\beq
\mathrm{N}'_{\textrm{enc}} (M, <r) = \mathrm{N}'_{\textrm{enc}} (10 M_\odot, <r)\left( \frac{m}{10 M_\odot} \right)^{-1.3} \; ,
\eeq 
where $\mathrm{N}'_{\textrm{enc}} (m, <r)$ indicates the number of stars of mass $M$ enclosed within a radius $r$ from Sgr A* per logarithmic bin. The mass function of the star cluster could be more top-heavy, although simulations of stellar dynamics show that a Kroupa mass function is still consistent for a star cluster near Sgr A* \citep{2009MNRAS.398..429L}. Assuming that the present moment is not a special time in the star formation history at the Galactic Center (i.e., the Galactic Center is not currently experiencing a starburst of massive stars), we can relate $N'(M, <r)$ to $\mathrm{N}'_{\textrm{obs}}(10 M_\odot, <r)$, the \emph{observed} number of enclosed stars in a logarithmic mass bin centered at $10 M_\odot$, 
\beq
\mathrm{N}'_{\textrm{enc}} (M, <r) \sim \mathrm{N}'_{\textrm{obs}}(10 M_\odot, <r) \frac{T(M)}{T(10 M_\odot)} \left( \frac{m}{10 M_\odot} \right)^{-1.3} \; ,
\eeq 
where $T(m)$ is the main sequence lifetime of a star of mass $m$. While red giants also contribute to the number of S-stars, the contribution of the giant phase lifetime to $T(m)$ is small compared to that of the main sequence lifetime, and thus are ignored in this approximation. $T(m)$ can be fitted by the formula \citep{lifetime1},
\beq
\log_{10} [T(m)/\mathrm{yr}] = 0.825 \log^2_{10} \left( \frac{m}{120 M_\odot} \right) + 6.43 \; .  
\eeq
This extra $T(M)/T(10 M_\odot)$ factor takes into account the fact that high mass stars do not live as long as low mass stars, and evolve into compact remnants that are unobservable.

The fact that the star S2, an S star of mass $\sim 10 M_\odot$ with an apocenter of $10^{-2}$ pc from Sgr A*, has been detected \citep{1998ApJ...509..678G,2016ApJ...830...17B,GRAVITYprecess} implies that 
\beq \label{eq:norm} 
\mathrm{N}'_{\textrm{enc}} (M_\odot, < 10^{-2} \; \textrm{pc}) \sim 20 \frac{T(M_\odot)}{T(10 M_\odot)}  \sim 8 \times 10^3 \; .
\eeq

The density of stars as a function of radius at the Galactic Center can be fitted by the Nuker model \citep{Lauer, Merritt2, 2016ApJ...821...44F,rho1, rho2, rho3},
\beq
\rho(r) \propto \left( \frac{r}{r_b} \right)^{- \gamma} \left[ 1 + \left(\frac{r}{r_b} \right)^\alpha \right]^{\frac{\gamma -\beta}{\alpha}} \; ,
\eeq
which consists of an inner power-law with index $\gamma$, an outer power law with index $\beta$, and a smooth transition region around the break radius, $r_b$ with $\alpha$ parameterizing the sharpness of the transition. Extrapolating to our regime of interests, the Nuker model reduces to its inner power law form,
\beq
\rho(r) \propto r^{-\gamma} \; ,
\eeq
where $\gamma = 1.75$ represents the Bahcall-Wolf cusp \citep{BWCusp}. While $\gamma$ has been measured for stars located further from Sgr A* \citep{rho1, Merritt2}, no measurement of $\gamma$ has been made for $r \lesssim 100 R_S$. Using equation (\ref{eq:norm}) for our normalization, Figure \ref{fig:Nenc} plots the enclosed number of $1 M_\odot$ stars per logarithmic bin, $N'_{\textrm{enc}} (M_\odot, <r)$ as a function of distance from Sgr A* for a variety of $\gamma$'s. We find that a star cluster at the center of the Milky Way Galaxy results up to tens of Solar mass stars orbiting within $100 R_S$. Such a star cluster might also host an even larger number of dwarf mass stars, which could dominate the occultation event rate, so Figure \ref{fig:Nenc} should be thought of as an order of magnitude estimate.

We note that the normalization equation (\ref{eq:norm}) overshoots the recent GRAVITY limit by a factor of $\sim 2$, and thus is most likely an overestimate \citep{GRAVITYprecess}. This might be the result of deviations from our assumptions of the mass function or star formation history close to Sgr A*. However, even with a normalization that is smaller by an order of magnitude, we still predict an appreciable number of stars of Solar mass and lower orbiting very close to Sgr A*. The event rate of Sgr A* occultations depends on this normalization, and thus in principle can be used to calibrate the currently unknown mass function and star formation history close to Sgr A*.

\subsection{Sizes of stellar occulters}
Occultations by larger stars provide a stronger signal, but owing to the mass function and the short main-sequence lifetimes of massive stars they represent much rarer events. On the other hand, while less massive stars produce smaller occultation signal, they might be numerous enough to dominate the number of detected occultations. For example, stars like Proxima Centauri, with a mass of $\sim 0.12 M_\odot$ and a radius of $\sim (1/6) R_\odot$ \citep{ProximaMass,ProximaRadii,ProximaCent}, will generate occultation signal that is weaker than occultations by Sun-like stars, but are $\sim 20$ times more numerous. Using the observational fit that the mass-radius relation of a star is approximately linear \citep{1999ApJ...514..725R}, the angular radius of an occulter at the Galactic Center distance of $\sim 8$kpc \citep{GCdistance} is,
\beq \label{eq:ocsize}
\theta_o^* (M) \approx 0.58 \times \left(\frac{M}{M_\odot}\right) \mu\mathrm{as} \; .
\eeq
Unlike the case with exoplanet occultation of stars, where the occultation signal is mainly observed as a flux reduction (transit depth) that scales with the area of the occulter, the SVLBI signal that we are considering include effects that scale with the diameter of the occulter. This is because the interferometry signal of a baseline, by the projection-slice theorem, is sensitive to the one dimensional size of the occulter along the baseline axis. 

Further, the metallicities of stars in the Galactic Center can be much greater than Solar values \citep{2009ApJ...691.1816N, 2018ApJ...855L...5D}. As the metallicity of a star determines the opacity of its atmosphere, there is a positive correlation between the sizes of stars and their metallicities \citep{2016ApJ...822...97H, 2019AJ....157...63K}. This means that occulters close to Sgr A* potentially possess significantly greater radii than that predicted by equation (\ref{eq:ocsize}), and thus have correspondingly larger occultation signals. 

\section{Enhancement of transit probabilities by general relativistic precession} \label{sec:precess}

\begin{figure}
\centering
\includegraphics[width=3in]{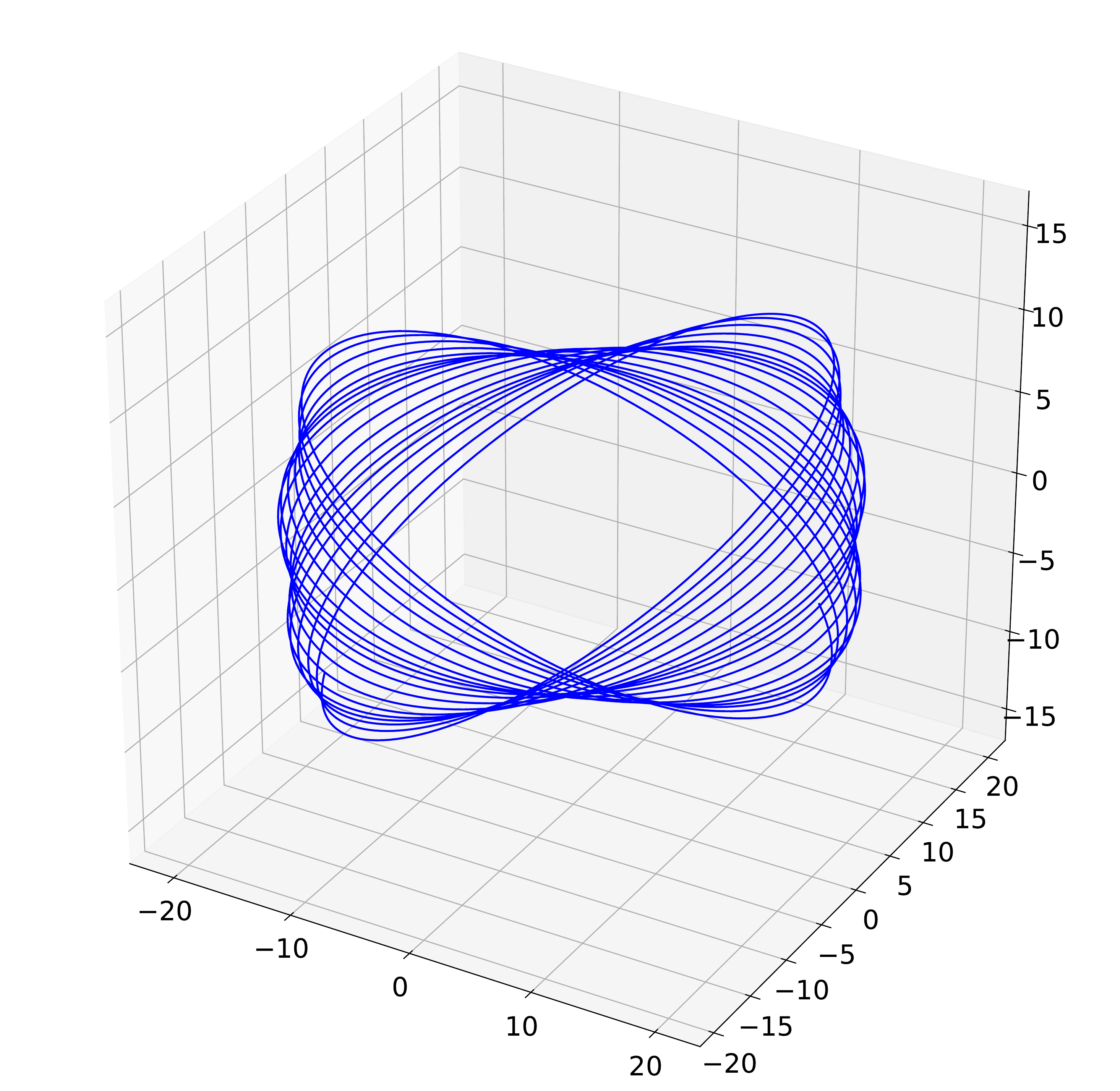} 
\caption{Geodesic integration of a star's orbit around Sgr A* with spin parameter $a=0.5$ and the black hole spin pointing in the vertical direction ($\sim 20$ orbits are shown over a time of $\sim5$ days, a fraction of $T_J$). The star orbits Sgr A* with an initial semi-major axis of $20R_S$ and initial inclination of $52$ degrees off the equatorial plane. The black hole is located at $(0,0,0)$. Axes labels are in Schwarzschild radius. The star's orbital node precesses around the spin axis, and causes the orbit to cover a large fraction of $4\pi$ solid angle after a precession period.}
\label{fig:geo}
\end{figure}

The probability for an object A to be seen in transit around another object B is,
\beq \label{eq:transit}
P_{\textrm{transit}} = \frac{ \Omega_O }{4 \pi } \; ,
\eeq
where $\Omega_O$ is the solid angle, as seen from B, covered by the track of object A as it orbits around B. For exoplanets, the probability of a planet orbiting a star of radius $R_*$ at orbital radius $a_p$ to be seen in transit is therefore given by $\sim R_*/a_p$. Using this formula for the probability of a Sun-like star to be seen in transit across the supermassive black hole Sgr A* gives a sizeable transit probability of $\sim 50 \%$ if the star is orbiting just beyond its tidal disruption radius at $\sim 10 R_S$.

However, due to general relativistic precessions, an object in close orbit around a supermassive black hole possesses an even greater transit probability. Obits inclined with respect to the black hole equatorial plane will have their orbital plane precess due to frame dragging and quadrupole precessions. These nodal precessions cause the star orbital path to fill a significant portion of the solid angle around the black hole. For a wide range of orbital parameters, $\Omega_O$ in equation (\ref{eq:transit}) becomes comparable to $4 \pi$. An illustrative case is shown in Figure \ref{fig:geo}, where we followed a numerical integration of the geodesic equation in the Kerr metric representing a star in an inclined orbit around Sgr A*. As shown in Figure \ref{fig:geo}, a timelike geodesic precesses and covers a large fraction of the entire $4\pi$ solid angle. 

\begin{figure}
\centering
\includegraphics[width=3.3in]{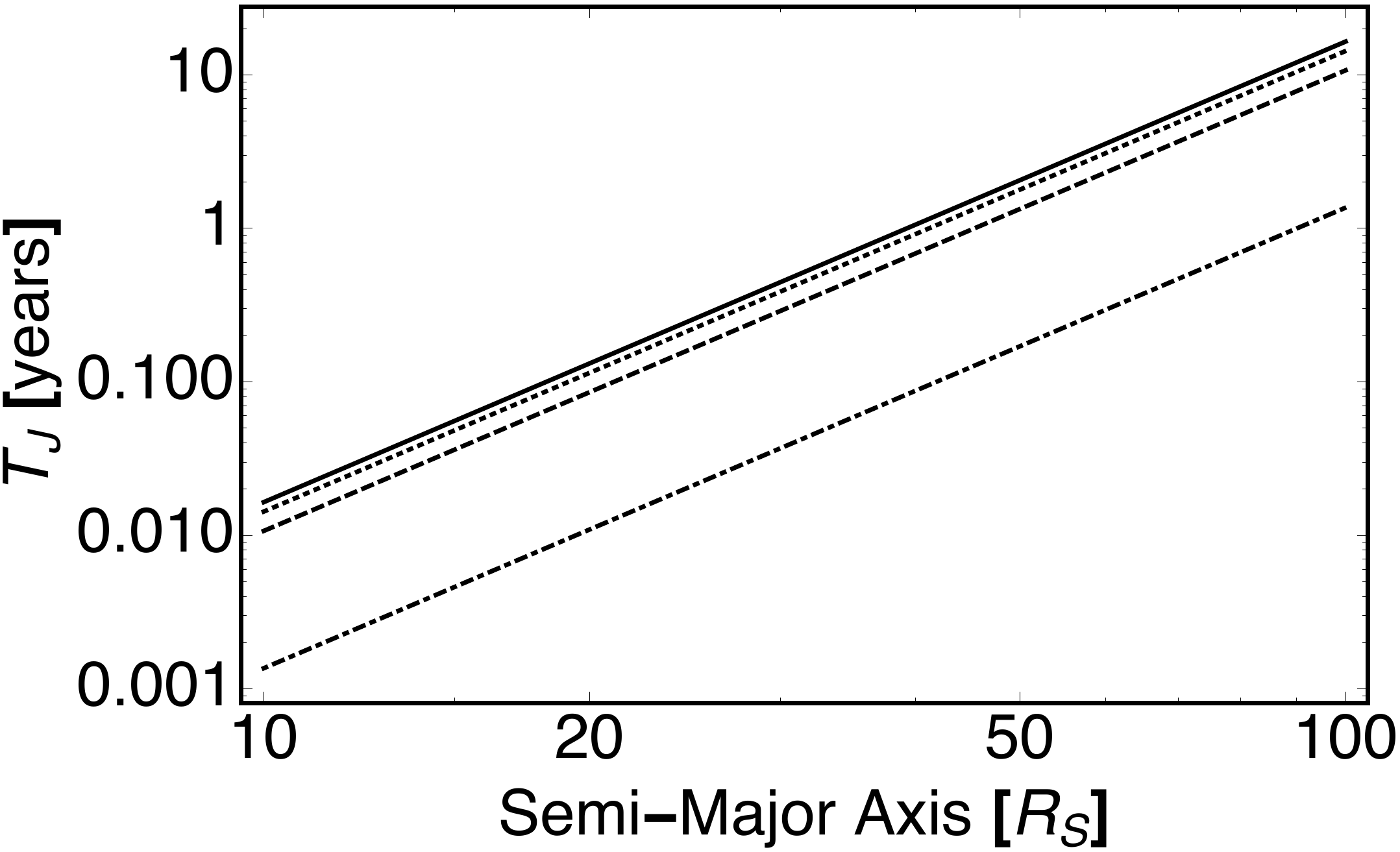} \\
\includegraphics[width=3.3in]{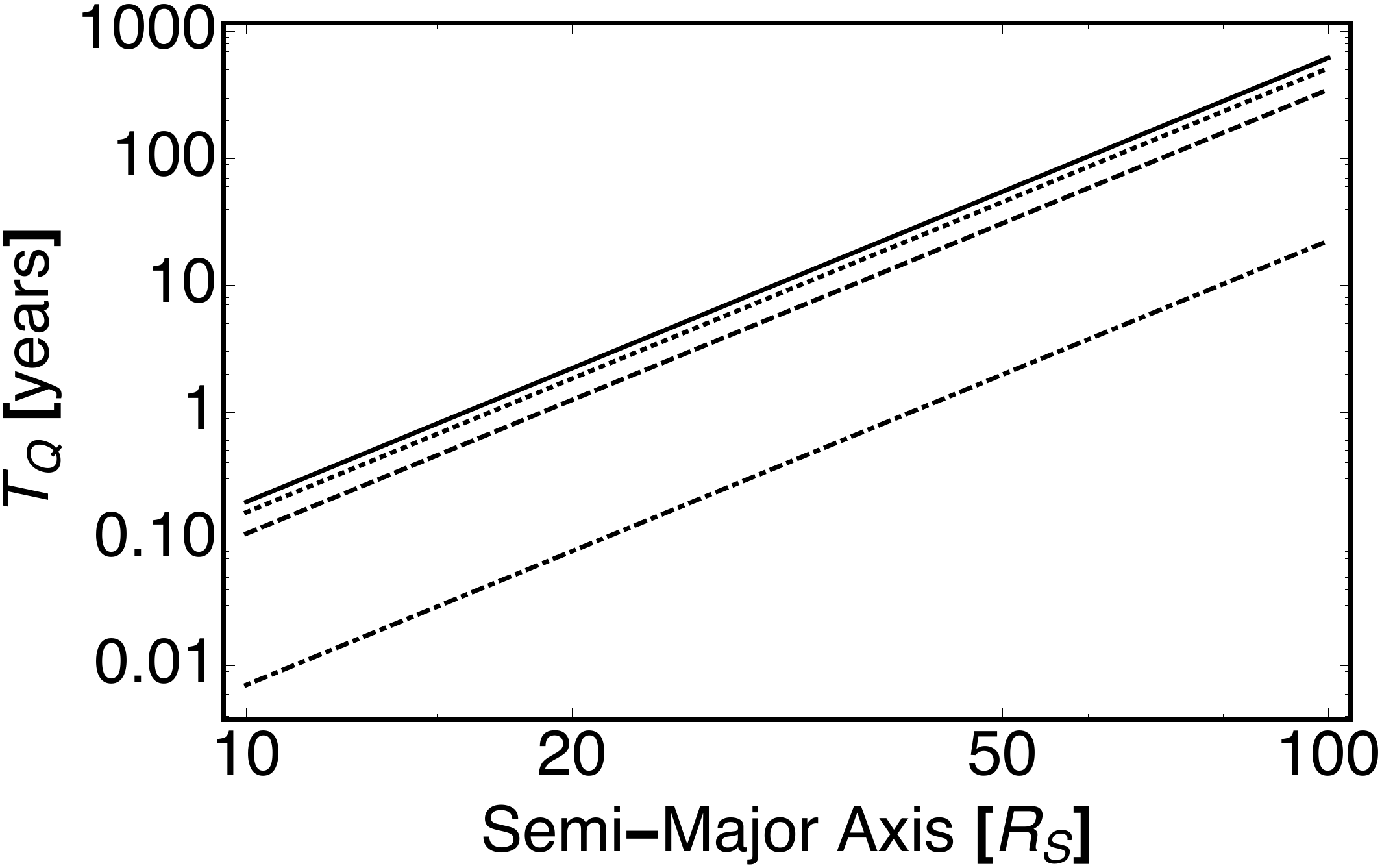}
\caption{Precession timescales for the frame dragging (\emph{top}) and quadrupolar (\emph{bottom}) precessions as a function of orbital semi-major axis around a black hole of spin $a=0.5$ with the mass of Sgr A* ($R_S \approx 10^{12}$cm). The different lines indicate orbits with eccentricities $0.9$ (solid), $0.5$ (dotted), $0.3$ (dashed), and $0$ (dot-dashed).}
\label{fig:Precess}
\end{figure}

If a star is located far from the black hole, however, the precession timescales are long, and thus will not increase the transit probability appreciably over a typical observational campaign. Therefore, unlike the exoplanet case, whether a star is seen in transit across Sgr A* or not is less a function of its inclination and position of orbital nodes, but rather whether the star orbits the black hole close enough to undergo significant precession. The frame dragging precession has a period of \citep{Merritt_book, 2013ApJ...777...57P},
\beq
T_J = \frac{P}{4 a} \left[ \frac{c^2 r (1-e^2)}{G M_{\textrm{BH}} } \right]^{3/2} \; ,
\eeq
where $P$ is the Keplerian period, $a$ is the black hole's spin, $e$ the orbital eccentricity, $r$ the orbital semi-major axis, and $M_\textrm{BH}$ the mass of the black hole. Figure \ref{fig:Precess} plots $T_J$ as a function of orbital semi-major axis around a $\sim 4\times 10^6 M_\odot$ black hole with $a=0.5$. The frame dragging precession timescale for stars orbiting such a black hole at semi-major axis $r \sim 50 R_S$ is $\sim 1$ year for non-eccentric orbits. If the orbit is highly eccentric ($e \ge 0.9$), this point is reached at $r \sim 100 R_S$.

A secondary nodal precession is supplied by the orbital interaction with the spacetime's quadrupole moment. The period for the quadrupole precession is given by \citep{Merritt_book, 2013ApJ...777...57P},
\beq
T_Q = \frac{P}{3 |q|} \left[ \frac{c^2 r (1-e^2)}{G M_{\textrm{BH}}} \right]^2 \; ,
\eeq
where $|q|$ is the spacetime's quadrupole moment. For a Kerr black hole, $q = -a^2$. Figure \ref{fig:Precess} plots $T_Q$ as a function of orbital semi-major axis around a $\sim 4\times 10^6 M_\odot$ black hole with $a=0.5$. In our regime of interest, the quadrupole precession is weaker than the frame dragging precession, and is only important for stars orbiting with semi-major axis $r \lesssim 50 R_S$ for eccentric orbits and $r \lesssim 20 R_S$ for non-eccentric orbits.

In addition to the relativistic precessions, other stars can also perturb the orbits of a star orbiting near Sgr A*. This effect would further increase the amount of solid angle covered by the star's orbit beyond its Keplerian estimate \citep{Merritt2}.

\section{Occultation model} \label{sec:model}

We employ for simplicity a crescent model for the black hole emission, obtained by subtracting a disc of radius $R_n$ from within a larger disc of radius $R_p$ in the image plane. The complex visibility of this model is given by \citep{2013MNRAS.434..765K},
\begin{align} \label{eq:emission}
V_e (u,v) =& 2 \pi I_0 \left[ \frac{R_p J_1(2\pi k R_p)}{2 \pi k} \right. \nonumber 
\\ &\left.  - e^{-2 \pi (a_1 u + b_1 v)} \frac{R_n J_1 (2 \pi k R_n) }{2 \pi k} \right] \; ,
\end{align}
where $I_0$ is the surface brightness, $a_1$ and $b_1$ are the horizontal and vertical offsets, respectively, of the inner disc from the center of the larger disc, $J_1(x)$ the Bessel function of the first kind, and 
\beq
k = \sqrt{u^2 + v^2} \; .
\eeq
When the the center of the smaller disc coincides with that of the larger disc, $a_1=b_2=0$, we will refer this model as the \emph{ring} model. Due to the small angular size of a typical occulter, its effects will only be seen in long baselines with lengths $>50 \textrm{G}\lambda$. The validity of using a crescent instead of a full GRMHD simulation in modeling our emission source in this regime is supported by the fact that while accretion astrophysics results in complex structures appearing in the visibility of short baselines, visibilities of long baselines where the occultation signal is most prominent are dominated by the clean signature of an emission ring \citep{Johnsoneaaz1310}. 

The occulter is modeled as a non-emitting disc of radius $R_o$ offset from the center of the larger emitting disc by $a_2$ in the horizontal direction and $b_2$ in the vertical direction in the image plane. This disc subtracts the flux from the emission model, so that the total visibility is given by
\beq \label{eq:signalV}
V (u, v) = V_e (u, v) - V_o(u,v) \; ,
\eeq
where 
\beq 
V_o (u,v) = F_U  \left[ H \left(R_o - \sqrt{(X-a_2)^2, (Y-b_2)^2} \right) \right] \; ,
\eeq
where $X$ and $Y$ are coordinates on the image plane, $H(x)$ the Heaviside function, and $F_U$ the Fourier transform where the domain is restricted to be within the emitting ring, $U$.

\section{VLBI signal} \label{sec:signal}

\begin{figure*}
\centering
\includegraphics[width=2in]{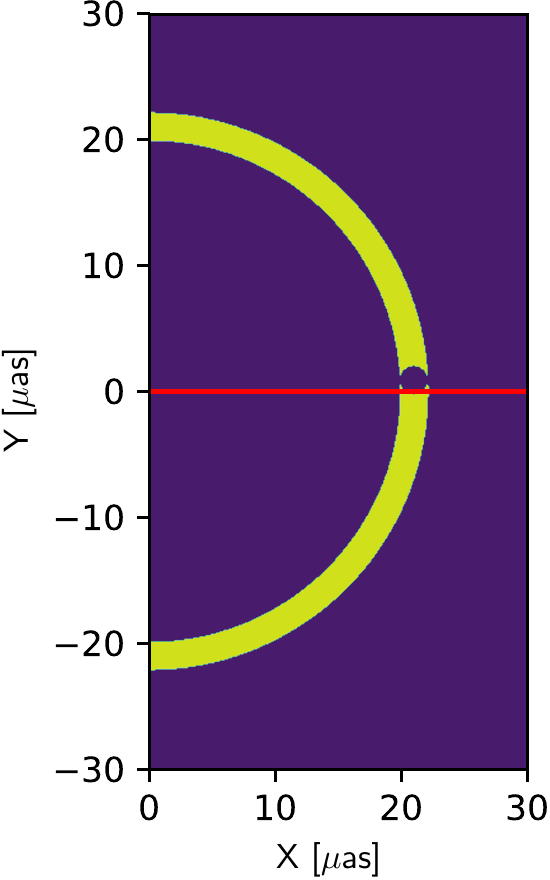} 
\includegraphics[width=2in]{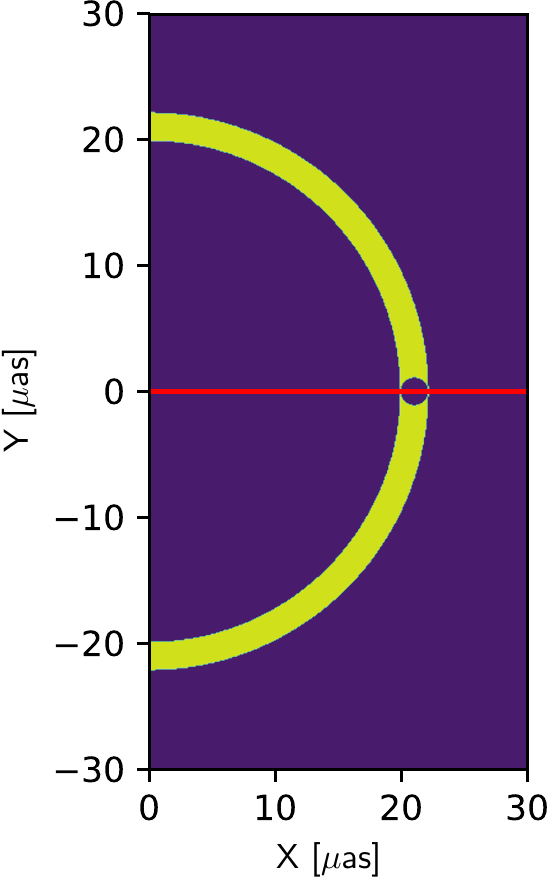} 
\includegraphics[width=2in]{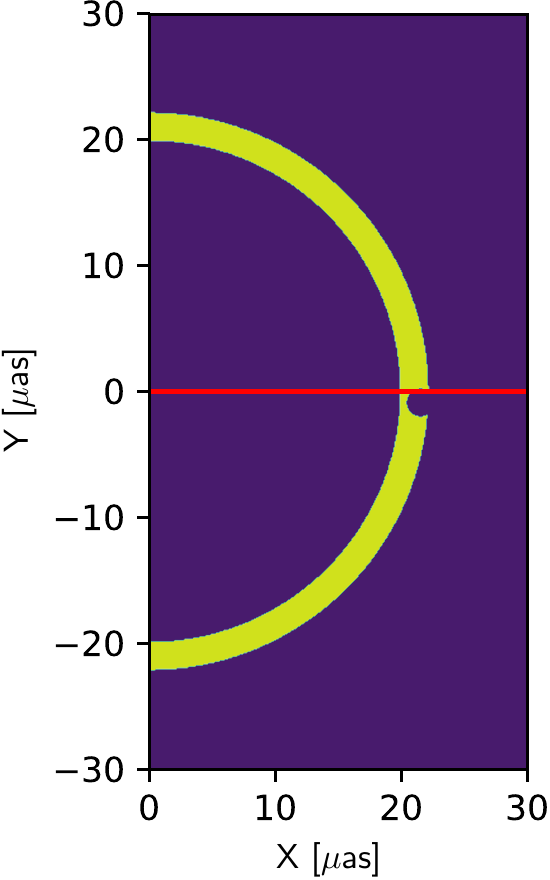}
\caption{The orbital motion of an occulter as it crosses the observational axis at $Y=0$ (red line) on the image plane. The emission model is given by equation (\ref{eq:emission}) with $a_1=b_1=0$, $R_1=22.1 \mu as$, $R_2=19.9 \mu as$, and the occulter is a star of radius $\sim 2 R_\odot$ orbiting at $100 R_S$. Due to the perojection-slice theorem, the signal as seen by a baseline along this observational axis depends only on the projection of the occulter along the observational axis (see text for details).}
\label{fig:orbit}
\end{figure*}

To gain an understanding of how the occultation signal, equation (\ref{eq:signalV}), is manifested in observations of an array with a limited coverage of $uv$-space, consider the case of a single SVLBI baseline. The orientation of this baseline with respect to Sgr A* defines an \emph{observational axis} on the image plane. When an occulter crosses the emission region, the VLBI signal along that axis will be modified by the presence of the occulter by the projection-slice theorem, 
\beq \label{eq:proj-slice}
F \circ P = F \circ S \; ,
\eeq 
where $F$ is the Fourier transform, $P$ the projection operator that acts on the image plane by an integral
\beq
P[f(x,y)] = \int_{-\infty}^{\infty} f(x,y) dy \; ,
\eeq
with $x$ and $y$ being the direction parallel and perpendicular, respectively, to the observational axis of the baseline, and $S$ is the slice operator, that acts on $uv$-space as 
\beq
S[f(u_x,v_y)] = f(u_x, 0) \; , 
\eeq
where $u_x$ and $u_y$ are the directions corresponding to $x$ and $y$, respectively, in $uv$-space. The right-hand-side of equation (\ref{eq:proj-slice}) is the signal measured by the SVLBI baseline, while the left-hand-side is simply an integral on the image plane. By the projection-slice theorem during an occultation event, the signal detected by a single VLBI baseline will be a the projection of the 'hole' created on the emission region due to the presence of an occulter to its observational axis. 

As a concrete example, Figure \ref{fig:orbit}, shows the orbital motion of an occulter of radius $\sim 2 R_\odot$ as it moves across an observational axis (aligned to $Y=0$ on the image plane) during a \emph{baseline crossing event}. That the occulter crosses an observational axis is in actuality not important, as by the projection-slice theorem, a baseline on this observational axis would still be able to detect this occulter even if the occulter is offset from said axis. In Figure \ref{fig:orbit}, before the circular occulter crosses the baseline, it is located above and to the left of the observational axis, while after the crossing it is located below and to the right of said axis. As the projection-slice theorem states that the baseline is blind to positional information along the vertical axis, the baseline just detects an occulter moving from the left to the right in the coordinates of Figure \ref{fig:orbit}.

\begin{figure*}
\centering
\includegraphics[width=6.5in]{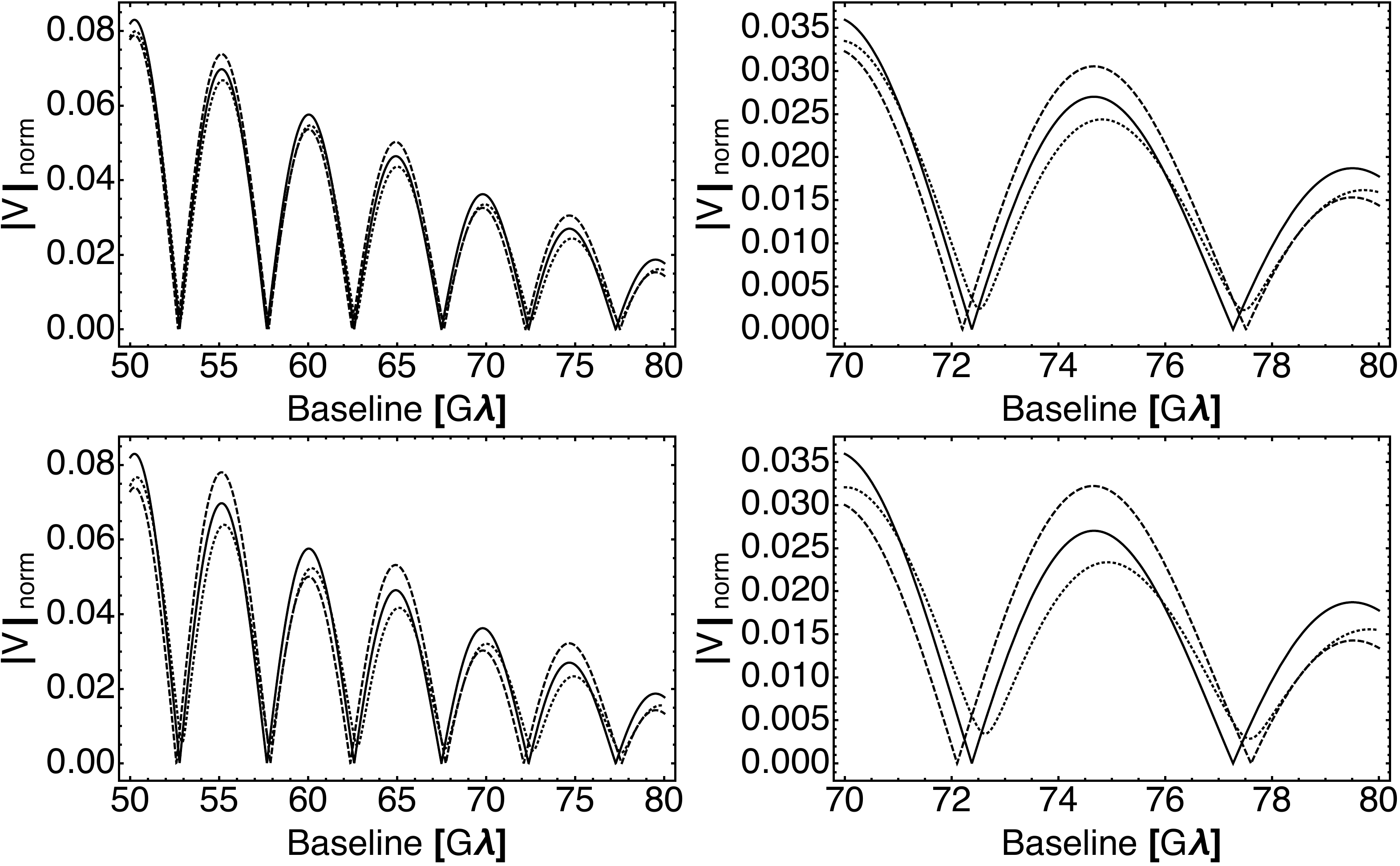}  
\caption{A snapshot of the normalized visibility amplitude in Fourier space for an occulter crossing the axis, as in the middle panel of Fig. \ref{fig:orbit} (dashed), when there is an occulter located along the $u_x=0$ line of the axis (dotted), and without an occulter (dashed) for a Sun-like star of radius $R_\odot = 7\times 10^{10}$cm (\emph{top}) and for a slightly larger occulter of radius $\sim 2 R_\odot$ (\emph{bottom}), both orbiting at $100 R_S$. The left panel shows the signal from $50-80$ G$\lambda$, while the right panel shows the same signal from $70-80$ G$\lambda$. The emission model is given by equation (\ref{eq:emission}) with $a_1=b_1=0$, $R_1=22.1 \mu as$, $R_2=19.9 \mu as$, and $I_0$ normalized so that the first peak at $u=v=0$ of the un-occulted model is unity. The occulter is placed at $a_2=21 \mu as$ and $b_2=0$ for the dashed lines, and at $a_2=0$ and $b_2=21 \mu as$ for the dotted lines.}
\label{fig:signal}
\end{figure*}

Figure \ref{fig:signal} shows a snapshot of the normalized visibility amplitude, $|V|_{\textrm{norm}}$, in the VLBI $uv$-space along such an axis when an occulter is in the middle of the axis. The amplitudes are normalized so that the first peak of the un-occulted model is unity. The presence of the occulter is imprinted on the $uv$-space as both a change in the visibility amplitudes, as well as a shift in the locations of the troughs and peaks of the signal. 

The situation for multiple SVLBI baselines is similar. If the baselines all lie across one observational axis, then multiple points in the $x$-axis of Figure \ref{fig:signal} can be sampled. This improves the detectability of an occultation event. If instead the baselines lie on multiple observational axes on the image plane, by the projection-slice theorem, the signal on each axes will be sliced onto each observational axes' corresponding $uv$ axes.

\subsection{Event timescales}

A baseline crossing event occurs over a timescale,
\beq
T_{c} = \frac{2 R_o}{v_o \sin{\theta_{\textrm{LOS}}}} \; ,
\eeq
where $\theta_{\textrm{LOS}}$ is the angle between the orbital velocity and the line-of-sight, while $v_o$ is the velocity of the occulter given by,
\beq \label{eq:vo}
v_o \approx \sqrt{GM \left( \frac{2}{r_i} - \frac{1}{r}  \right) } \; ,
\eeq
where $r_i$ is the instantaneous orbital distance from Sgr A*, and $r$ is again the orbital semi-major axis. The approximate sign in equation (\ref{eq:vo}) refers to the fact that we are neglecting higher order relativistic effects. 

For a circular orbit, $v_0 \approx \sqrt{GM/r}$, and
\beq \label{eq:baselinetime1}
T_c \approx 1 \; \textrm{min} \times \frac{R_o}{R_\odot} \times \left( \frac{100 R_S}{r} \right)^{1/2} \times \sin^{-1} \theta_{\textrm{LOS}} \; .
\eeq
This timescale can be significantly different for eccentric orbits. For example, an orbit at apnegricon with $r_i=(1+e)r$, $r=100 R_S$, and $e=0.7$ possesses a baseline crossing time of 
\beq \label{eq:baselinetime2}
T_c \approx 2.5 \; \textrm{min} \times \frac{R_o}{R_\odot} \times \sin^{-1} \theta_{\textrm{LOS}} \; .
\eeq
Further, due to the geometric factor $\sin{\theta_{\textrm{LOS}}}$, this timescale can be very long for eccentric orbits with a geometry where the orbital velocity is mostly along the line-of-sight. Related to the baseline crossing time is the \emph{transit time}, defined as the time it takes the occulter to travel across a single section of the emitting region. For an occulter crossing normal to an emitting ring, this is the time it takes to cross the width of the ring. The transit time sets the timescale for the duration over which an occultation event can be detected for. If $L_{\textrm{limb}}$ is the characteristic size of a limb of the emitting region, the transit time is given by,
\beq \label{eq:transittime}
T_t = \frac{L_{\textrm{limb}}}{v_o \sin{\theta_{\textrm{LOS}}}} \; ,
\eeq
if the limb is larger than the occulter. If the occulter is larger than the limb ($2R_o > L_{\textrm{limb}}$), then the transit time is equal to the baseline crossing time,
\beq
T_t = T_c \; .
\eeq
While $L_{\textrm{limb}}$ takes a large range of values that can depend on the geometry of the crossing or, in the case of a crescent emission region, which side of the crescent the occulter passes through, its minimum value is of order minutes. While we do not know the future capabilities of SVLBIs, we note that the EHT scans are minutes long \citep{PaperII}.

Finally, the timescale for the occulter to cross the entire emission region (i.e., the entire ring or crescent), is 
\beq
T_{e} \approx \frac{\sqrt{27} R_S}{v_o \sin{\theta_{\textrm{LOS}}}} \; ,
\eeq
though this duration can be significantly shorter if the occulter crosses near grazing incidence of the emitting region. If an occulter is detected to be crossing the observational axis of a baseline, it will cross other observational axes within time $T_e$, which for circular orbits, 
\beq 
T_{e} \approx 0.8 \; \textrm{hr} \times \left( \frac{100 R_S}{r} \right)^{1/2} \; .
\eeq
For eccentric orbits, as we are now integrating over an appreciable fraction of the orbit, we adopt the average orbital speed,
\begin{align}
\bar{v}_o &= \frac{4 a E (e)}{T_K} \nonumber
\\ &= \sqrt \frac{G M}{r} \left( 1-\frac{1}{4}e^2 - \frac{3}{64}e^2 - \frac{5}{256}e^6  - \ldots  \right) \; ,
\end{align}
where $T_K$ is the Keplerian period and $E(m)$ is the complete elliptical integral of the second kind. On average, a higher eccentricity will increase the time taken for the occulter to cross the emission region. As before, there are special orbits where the geometry is just right for the $\sin{\theta_{\textrm{LOS}}}$ term to cause an extremely long $T_e$. However, these orbits require specific geometries and are rare. Neglecting these rare orbits as well as unbound orbits (where $T_e \rightarrow \infty$), the maximum timescale for the occulter to cross the emission region is,
\beq \label{eq:Temax}
T_{e, \textrm{max}} \approx 1.3 \; \textrm{hr} \times \left( \frac{100 R_S}{r} \right)^{1/2} \; .
\eeq
Beyond this timescale, one would have to wait for at least an orbital time before the next occultation event by the same occulter.

\subsection{Methods for identifying false positives}

\begin{figure}
\centering
\includegraphics[width=3in]{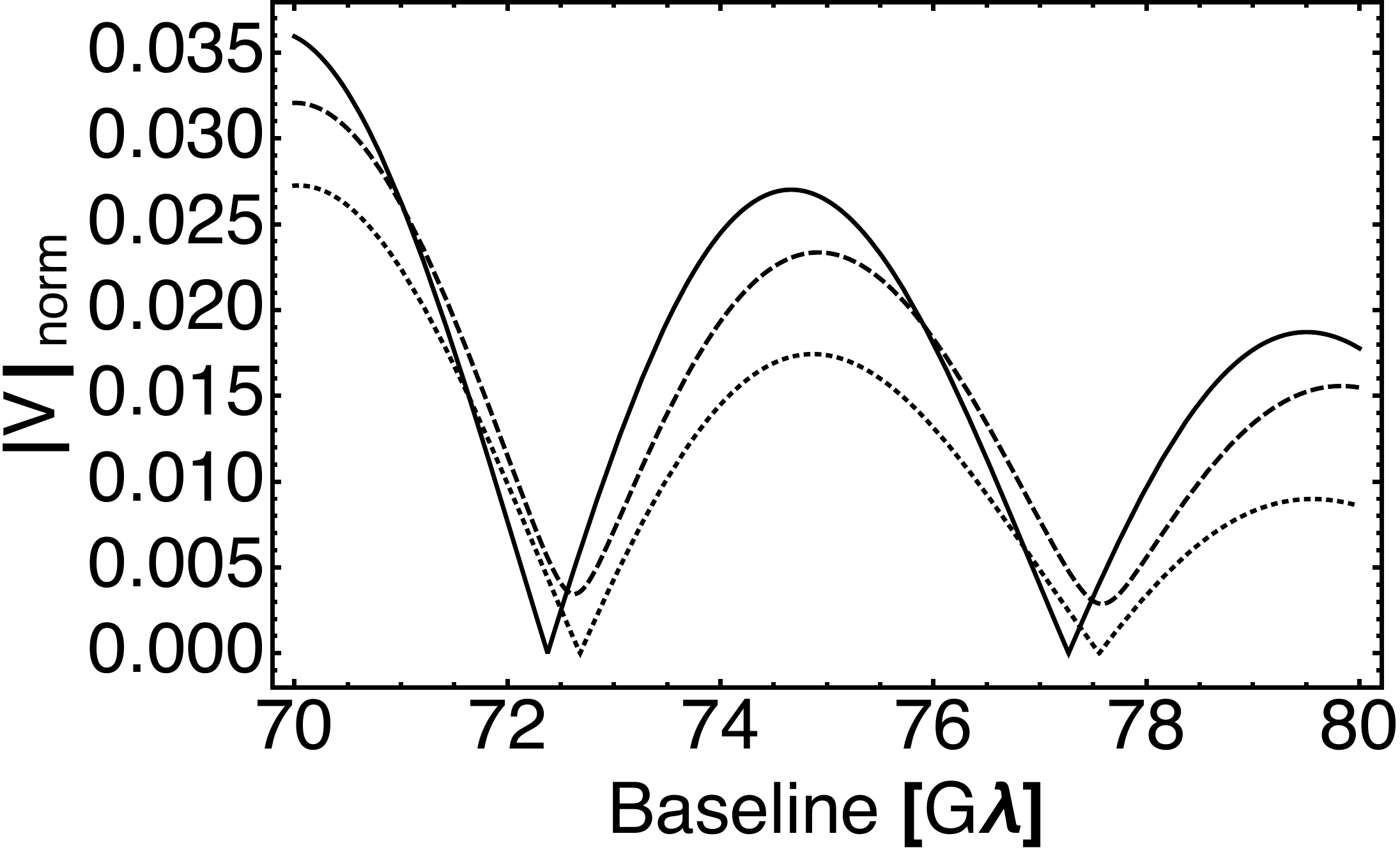} 
\caption{A snapshot of the normalized visibility amplitude in Fourier space for an occulter crossing the observational axis (dashed), without an occulter (solid), and a false positive (dotted) produced by reducing the thickness of the emission ring. For an instance of time, a false positive produces the same qualitative signal as an occulter, i.e., changing the amplitude and positions of the troughs and peaks of the visibility amplitude.}
\label{fig:mimicker}
\end{figure}

As discussed in the previous subsections, multiple baselines along the same observational axis provide more sampling points of the signal during an occultation event, while multiple baselines on different observational axes provide the opportunity for multiple baseline transit events to be observed by different baselines, separated by time at most $T_{e, \textrm{max}}$. However, another important reason to require multiple SVLBI baselines is to break the degeneracy between an occultation event and inherent time-variations in the emission profile.

Along one observational axis, the qualitative signal due to an occulter can be mimicked by, for example, a transient thinning of the emission profile. We plot one such a false positive in Figure \ref{fig:mimicker}. As the orbital motion of the occulter produces a specific time variability in the signal, breaking the degeneracy between an occultation event and a false signal would be possible with one baseline if one has enough time resolution to resolve the transit. However, it could be challenging to use this method to characterize the shortest transit events, as those lasts for $T_t=T_e \sim$ minutes. 

If multiple baselines along the same observation axis are available, but not enough time resolution is achieved to resolve the transit, a mimicking event can be distinguished from an occulter through model fitting. Only very specific changes to the emission profile, ones that create circular holes in the emission profile, can mimic the full $uv$-signature of an occulter. Signals generated by thinning the emission ring along the observational axis, for example, can produce the same salient features as an occulter crossing the axis: a reduction in the overall amplitude and pushing the peaks and troughs of the signal to higher baselines (c.f. Figure \ref{fig:mimicker}). However, such an event cannot produce the same ratio of the offsets of the locations of the peaks/troughs to the reduction in amplitude as a genuine occultation event. 

If multiple baselines along different observational axes are available, one can again rule out false signals even if not enough time resolution is achieved in units of $T_t$. As implied by the projection-slice theorem, the signal of an occulter appears on all baselines through the Fourier transform of their projection on each baselines' observational axis. Suppose a candidate occultation event is detected at a particular baseline. If the occultation event is genuine, every other axis must, at the same instance, also have their signal modified by the amount demanded by the projection-slice theorem. An event created by generic transient variations in the emission region (e.g., changes in the thickness in a part of the emission region) will not create such a signal across multiple observational axes. As in the previous method, only very specific events, such as inherent variations that create circular holes in the emission profile, can pass this test.

Unlike an event resulting from changes in the emission profile, an occulter has to move along a straight line in the image plane. As such, when an occulter candidate event is detected, its time variation must behave as demanded by a linear motion. This fact allows us to further reject false positives using baselines on multiple observational axes, even if we do not have enough time resolution on the scale of a transit time, $T_t$. If a single limb crossing is detected, within time $T_{e, \textrm{max}}$, there will potentially be a secondary limb crossing event at a different location on the image plane, due to the occulter moving across the emission region. The detection of a candidate secondary crossing event within time $T_{e, \textrm{max}}$ increases the likelihood that both events are genuine. Conversely, a detection of a secondary event long after $T_{e, \textrm{max}}$ means that the two events are likely not occultation events. The tradeoff of this method is that it will mischaracterize genuine occultation events that cross the emission profile tangentially along chords far from the center of the emission region, as those occulters will only have a single limb crossing event. This method will also mischaracterize occulters along a hyperbolic orbit, or those on the rare orbits with geometry that allow them to take more than $T_{e, \textrm{max}}$ to cross the emission profile.

Suppose now that we have both multiple baselines on multiple observational axes and enough time resolution to resolve a transit time. Because occulters have to move along a straight line in the image plane, in a genuine occultation event, every single baselines must detect signal that are geometrically consistent with the motion of the occulter. For example, if the occulter's orbit is moving parallel to an observational axis, baselines on that axis must all detect the same motion moving in the same direction, and baselines on axes rotated with respect to the first must detect the same motion projected on their axes. If a secondary limb crossing event is also detected, the location of the secondary limb crossing can also be checked for consistency with the previously deduced motion of the occulter.

Finally, an occultation event will repeat over an orbital period until general relativistic precession brings the orbit out of transit. As such, the likelihood of an occultation event is improved through the detection of correlated occultation signals with a periodicity comparable to an orbital period ($\sim 4$ days for an orbit with a semi-major axis of $100 R_S$). As false positives might also exhibit variations with a timescale comparable to the orbital period, this method is unreliable if used alone, and is best used as a secondary method to strengthen the confirmation of an occultation event that passed the other checks.

\subsection{A case study with three baselines}

\begin{figure*}
\centering
\includegraphics[width=5in]{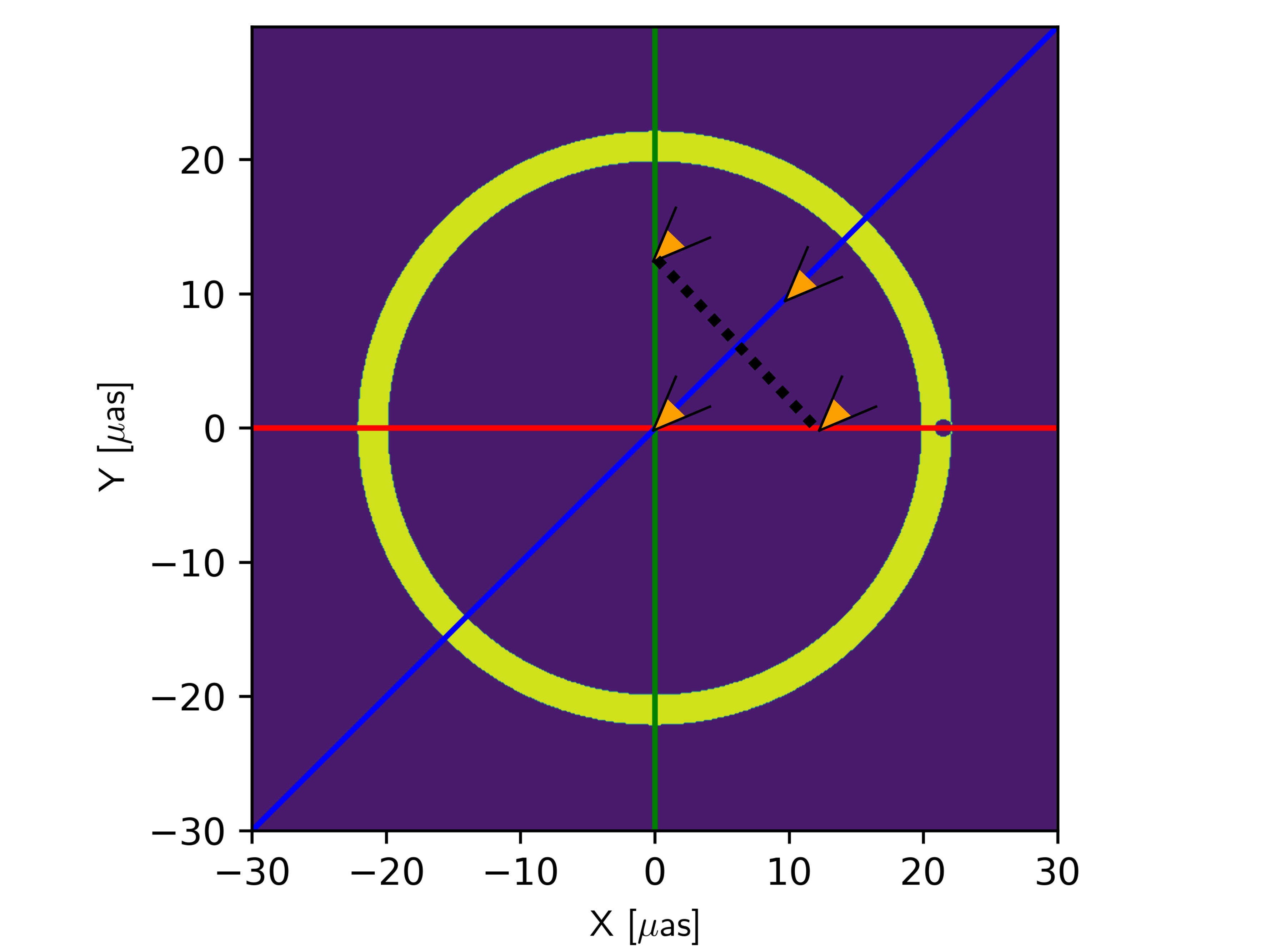} \\ 
~\\
\includegraphics[width=3in]{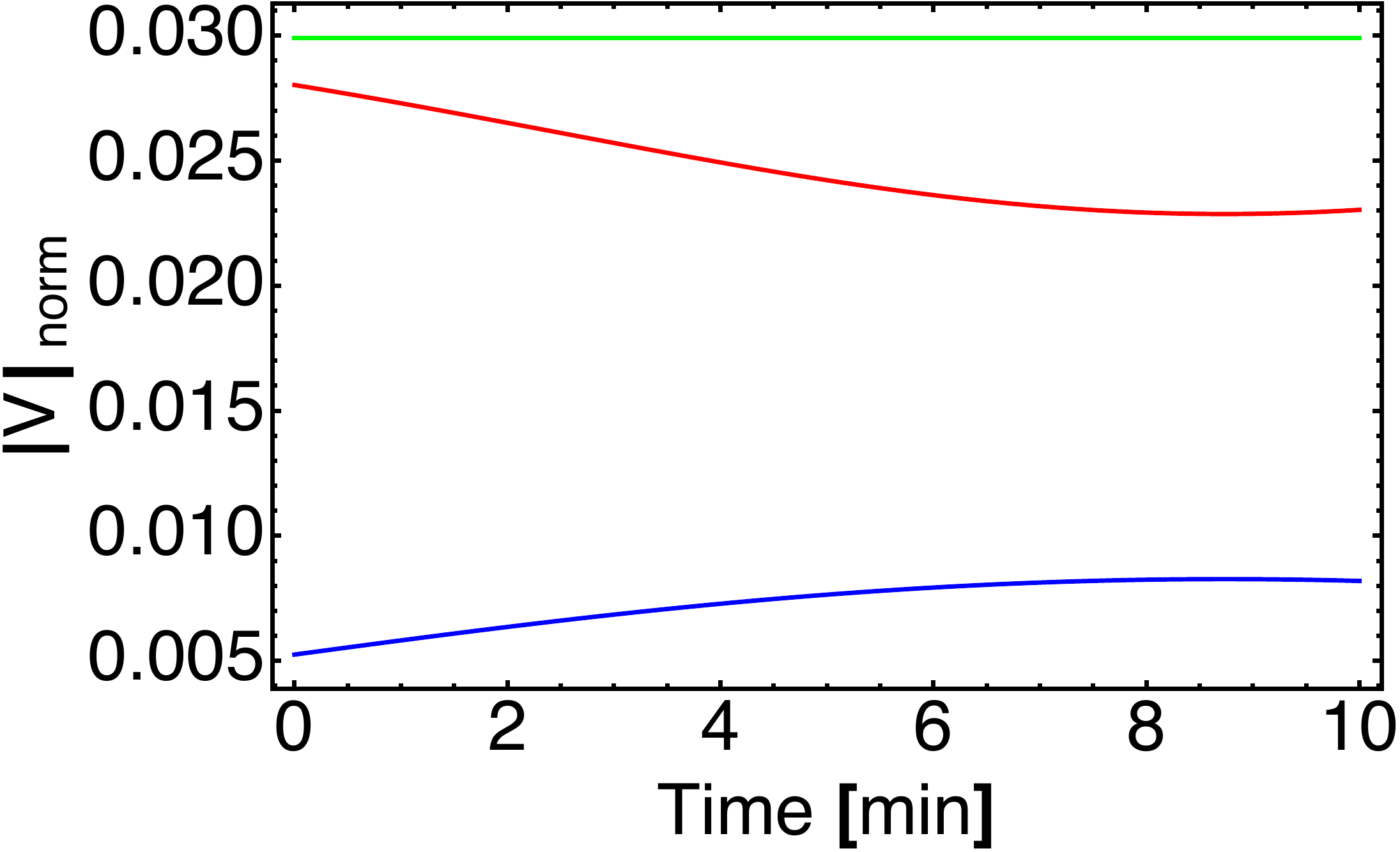} 
\includegraphics[width=2.83in]{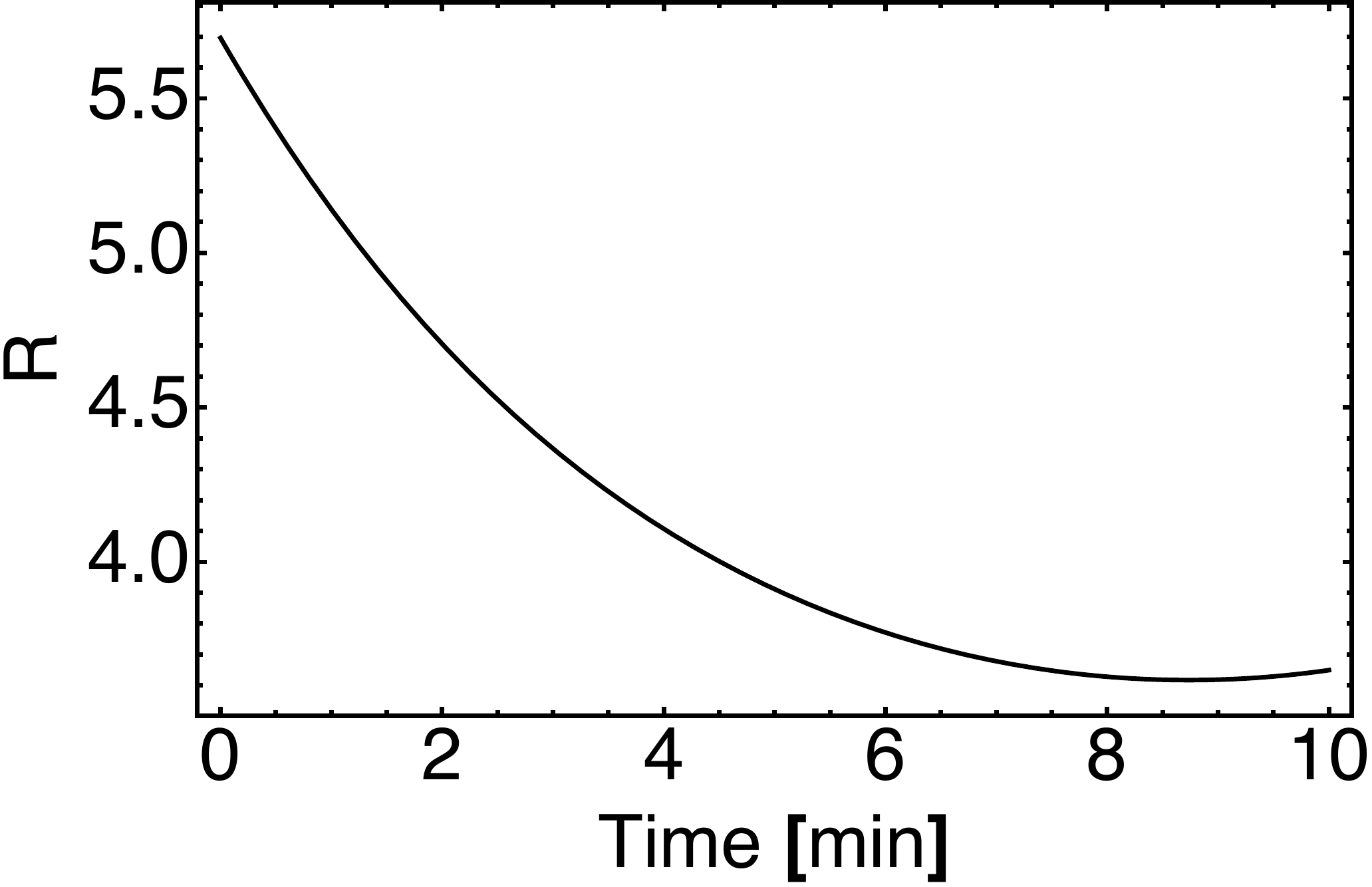} 
\caption{The emission ring and an occulter of Sun-like radius, $R_\odot = 7 \times 10^{10}$ cm, at coordinates $(X,Y)=(21.5,0)$ with the configuration of four SVLBI satellites (orange triangles) superimposed over it (\emph{top}). The orientation of three baselines generated by this configuration are plotted in green, blue, and red. For our configuration of satellites, the satellite at $(0,0)$ is separated from all the other satellites by a distance of $75$ G$\lambda$, thus giving us a coverage at $75$ G$\lambda$ in the $uv$-plane for all three baseline axes. This SVLBI configuration also possesses other baselines (e.g., the dotted black line) that can be used to further improve the detection. The SVLBI signal at $75$ G$\lambda$ as a function of time when the occulter moves along the negative negative $Y$ axis for the green, blue, and red baselines (\emph{bottom left}), as well as the ratio of said signal detected by the red and blue baselines (\emph{bottom right}). This occulter traverses across the limb at a timescale of $10$ minutes.}
\label{fig:3baselines}
\end{figure*}

In this subsection we explore the effect of having a sparse baseline coverage on the ability of an SVLBI experiment to distinguish a genuine occultation event from a false signal through a case study of an SVLBI with three baselines. Suppose we have an occultation event where the occulter is a Sun-like star detected by four stations configured in a manner plotted in the top figure of Figure \ref{fig:3baselines}. Analogous calculations for a different choice of station configuration can be made analogously and thus this choice of configuration can be made with no loss of generality, with the caveat that the stations are located far enough from each other for the occultation signature to be detectable. Further, we will limit our analysis to only three baselines, the red, green, and blue baselines plotted in Figure \ref{fig:3baselines}. While the other baselines (e.g., the dotted line in Figure \ref{fig:3baselines}) can be used to further improve the detection, we limit ourselves to only a subset of the baselines to demonstrate that a genuine occultation event can be distinguished even at times that, due to the orbital motion of the satellites, some of the baselines are too short to be able to detect an occultation signal.

With such a sparse array, we need to leverage the time variability of the signal to screen for genuine occultation events. Even in the chance that random fluctuations in the accretion flow can mimic the signal on all three baselines at a particular time, it is exceedingly unlikely for them to mimic the time variability of a genuine occultation event. This can be understood through the projection-slice theorem: when the occulter moves in image space (e.g., along the negative $Y$ axis in Figure \ref{fig:3baselines}), the portion of the image that has their intensity reduced due to the presence of the occulter moves accordingly on the image plane, and thus its projection on the baseline axes also moves. When Fourier transformed, this generates a predictable signal on each baseline. Assuming the simple model that the occulter moves at a constant velocity in a straight across the image plane, this signal can then be searched for using standard model fitting algorithms.

In the bottom figures of Figure \ref{fig:3baselines}, we plot the amplitude detected on the three baselines (all of them at $75$ G$\lambda$) as well as $R$, the ratio of the signal detected by the red and blue baselines. Because the occulter motion is perpendicular to the green baseline, the ratios between the signal detected by the green and the other baselines are trivial. While the ratios at a particular time can be mimicked by random fluctuations in the accretion flow, the time evolution shown in these plots are characteristic to a circular dark spot in the image plane moving with a constant velocity. 

There is the possibility for a cold spot orbiting along the accretion flow that can mimic the time variability of an occultation signal for short times. However, a cold spot is carried along the accretion flow, in contrast to an occulter that moves along its own orbit. Thus, a cold spot will not move along a straight line in the image plane except for a limited amount of time. If the directionality of the accretion flow is known (e.g., via Doppler beaming), a cold spot will also have a motion that is on average along said direction, in contrast to an occulter that can move unconstrained on the image plane. Further, as detailed in the previous subsection, an occulter can produce two occultation events within a short timescale if its motion strikes a chord across the emission region. This behavior is not present for cold spots. 

\section{Conclusions} \label{sec:conclusion}

We have shown that the stellar cluster around Sgr A* potentially results in an appreciable population of stars down to the tidal disruption radius at $\sim 10 R_S$. These stars are close enough to the supermassive black hole to facilitate occultation events, and modify the emission signal of Sgr A* as seen from an SVLBI. We have shown how the transit probabilities of stars orbiting Sgr A* are greatly enhanced by general relativistic precession. Even for modest Sgr A* spins of $a\sim0.5$, the precession timescales are short for stars orbiting with semi-major axis $r \lesssim 100 R_S$, especially on eccentric orbits.

Further, we computed how an occultation affects the SVLBI signal during a transit event. We found that, due to the size of a typical stellar occulter, the occultation signal is prominent at long baselines $\gtrsim 50 \textrm{G} \lambda$. The most generic signature of an occultation signal is the modulations of the amplitude, as well as the locations of the peaks and troughs of the signal. Having baselines that are oriented along different axes in the image plane is the best way to distinguish genuine occultation events from transient variations in the emission profile of the region around the black hole. 

\section{Acknowledgements}
The authors would like to thank Dimitrios Psaltis and the anonymous referee for useful comments. AL was supported in part by the Black Hole Initiative at Harvard University, which is funded by JTF and GBMF grants.

\bibliography{BibFile.bib}

\end{document}